\documentclass[%
 reprint,
superscriptaddress,
 amsmath,amssymb,
 aps,
]{revtex4-2}

\usepackage{graphicx}
\usepackage{dcolumn}
\usepackage{bm}
\usepackage{xr-hyper}
\usepackage{hyperref}
\hypersetup{
    colorlinks=true,
    linkcolor=blue,
    citecolor=blue,
    urlcolor=blue,
    }
\usepackage[mathlines]{lineno}
\usepackage{mathtools}
\usepackage{epstopdf}
\usepackage{color}
\usepackage{physics}
\usepackage{placeins}
\usepackage{multirow}
\usepackage{url}
\usepackage{xcolor}
\usepackage{amsfonts}
\usepackage{amsmath}
\usepackage{amssymb}
\usepackage{tikz}
\usepackage{dirtytalk}

\makeatletter
\newcommand*{\addFileDependency}[1]{
\typeout{(#1)}
\@addtofilelist{#1}
\IfFileExists{#1}{}{\typeout{No file #1.}}
}\makeatother

\newcommand*{\myexternaldocument}[1]{%
\externaldocument{#1}%
\addFileDependency{#1.tex}%
\addFileDependency{#1.aux}%
}
\newcommand{\HMMone}[1][]{in the main text.}
\myexternaldocument{HMMone}

\begin{document}

\preprint{APS/123-QED}

\title{Non-monotonic temperature dependence of light-matter interaction in hyperbolic metamaterial due to interplay of electron-phonon scattering}

\author{Amitrajit Nag}
\email{amitrajitnag@iisc.ac.in}
 \affiliation{
 Indian Institute of Science, C.V. Raman Road, Bangalore, India, 560012}
 \author{Jaydeep K. Basu}
 \email{basu@iisc.ac.in}
 \affiliation{
 Indian Institute of Science, C.V. Raman Road, Bangalore, India, 560012}
 
\begin{abstract}
Hyperbolic metamaterials (HMM) are artificially engineered materials that are congenial for light-matter interaction studies and nanophotonic applications with the hyperbolic dispersion of light propagating through them, which offers a large photonic density of states. We have explored HMM's broadband cavity-like modes and ultrasmall mode volumes, even though the system has lossy plasmonic constituents. The light-matter interaction properties of plasmonic materials strongly depend on different internal damping mechanisms. Temperature is a macroscopic parameter that controls these internal mechanisms and is reflected in their corresponding interaction behaviors. 
In this work, we investigated the light-matter interaction properties of the HMM system with temperature. We studied the HMM system weakly coupled to quantum emitters. This weakly coupled system shows a non-monotonicity in its broadband absorption and the emission from near-field coupling with quantum emitters. This is determined by the interplay between the electron-phonon and the phonon-phonon scatterings occurring in the metal nanowire array, effectively providing the damping with temperature. Theoretically, we confirmed the increased presence of the phonon-phonon scattering in nanowires compared to bulk metals, which plays an instrumental role in the observed light-matter interaction effects. This study could efficiently predict the use of the HMM in optics and photonics applications, with precise tuning and availability of control parameters with temperature. Also, this study could help identify the effect of increased phonon-phonon scattering in nanostructures and explore the possibility of quantifying and applying it by optical measurements.  \end{abstract}

\maketitle


\section{\label{sec:intro} Introduction}





Plasmonic materials offer strong confinement of electromagnetic fields into subwavelength volumes, which is a necessary key to efficient nanophotonic applications \cite{plasmonics_appl1, plasmonics_appl2, plasmonics_appl3, plasmonics_appl4, plasmonicNW_appl_nonliear}. The physical mechanism behind the confinement of the irradiating light to a volume much smaller than the diffraction limit in metal is the excitation of surface plasmon polaritons (SPPs)—partially coherent oscillations of free electrons in the material resonantly driven by the external electromagnetic waves \cite{Light_Concentration_Plasmonic}. The hybrid electron-photon nature of SPP modes offers opportunities to miniaturize optical networks and their hybridization with electronic circuits. This helps design functional optical metamaterials for the infrared and visible spectral ranges, integrating with resonant optical elements where the strong light confinement effect would strengthen the functionality   \cite{SPP_subwavelength_review, ENZ_MetaMaterial_NatPhot}. The detrimental effect comes from the faster dephasing and decay times of surface plasmons, which lie in the femtosecond time scale for radiative damping and generate energetic or \lq{hot}\rq carriers \cite{hotcarrier_Plasmonics_review, hotcarrier_Plasmonics, hotcarrier_Plasmonics_review2}. This \lq{hot}\rq carrier generation follows the Landau damping, a quantum mechanical process for plasmons to dissipate non-radiatively \cite{LandauDamp_theory, LandauDamping}. Interestingly, Landau damping can modify the dynamics of a quantum emitter coupled to the SPP \cite{LandauDampTheory_withQD}. Fast dephasing and decay times of plasmons translate into the broader resonant features of the plasmonic nanostructures, associated with the material losses of metals and local heating in the device via absorption.
\begin{figure}[b]
    \centering
    \includegraphics[width=0.48\textwidth]{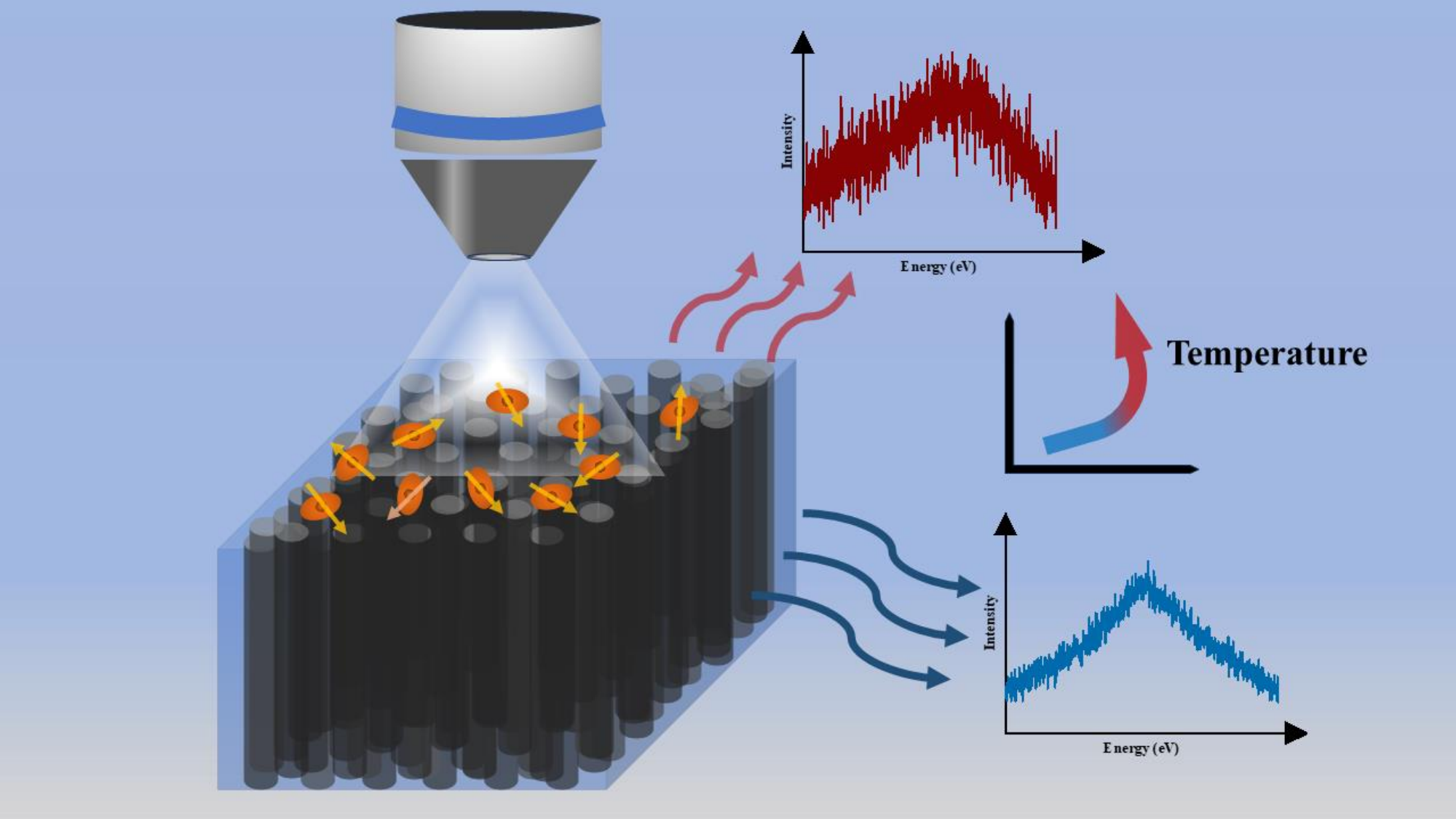}
    \caption{Schematic of the model HMM-quantum emitter system under investigation for light-matter interactions from cryogenic temperature to room temperature, showing the effect of increased damping, reflected as the noise in the schematics of emission spectra.}
    \label{fig:HMM_schematic_paper2}
\end{figure}
\\
SPP properties of plasmonic metamaterials are also malleable to ambient temperature due to the constituent metallic nanostructures, which provide the plasmon damping \cite{Plasmon_TempDepend_1, Plasmon_TempDepend_2, Plasmon_TempDepend_3, TempDepend_Ag_nanoparticle}.
Hyperbolic Metamaterial (HMM) is an artificially engineered material made of metal-dielectric composite \cite{HMM_review_paper2, HMM_rev2_paper2, HMM_rev3_paper2, HMM_rev4_paper2, HMM_rev5_paper2} to show hyperbolic dispersion of available photonic states after an optical topological transition (OTT) in the frequency space \cite{OTT1_paper2, OTT2_paper2, OTT3_paper2}. HMM has been successful in leaving an impression in various aspects of engineering applications like super-resolution imaging and microscopy \cite{HMM_hyperlens, HMM_superresolution}, bio-sensing \cite{HMM_biosensing, HMM_biosensing2}, energy harvesting \cite{HMM_energyHarvest}, applications as sensors due to its broadband absorption \cite{HMM_sensorAbsorber}, etc. HMM can strongly modify the available density of states, leveraging the Purcell enhancement significantly \cite{jacob2012broadband, HMM_RKY2}, and could show strong coupling between quantum emitters and HMM modes \cite{HMM_SRK, HMM_Harsha}. HMM's payoffs qualify it as an effective metamaterial system for a myriad of light-matter interaction studies and applications. In this regard, controlling the plasmon damping is essential for optimal use of such metamaterials. Detrimental effects of plasmon dissipation through dephasing and decays may get modified by the choice of size and shape of constituent metal nanoparticles, and the geometric effects, like the array in the metamaterial design \cite{AgNW_geometryDepend, Plasmon_geometric_1, Plasmon_geometricshape_2, Plasmon_array_1, Plasmon_array_2, Plasmon_array_3, Plasmon_array_4}.\\
In this work, we aimed to explore the dependency of light-matter interactions on the plasmonic damping for the silver nanowire (AgNW) array-HMM and how that can be controlled by tuning the damping properties. Temperature plays a significant role in determining the damping properties. Plasmonic damping mainly occurs due to the metallic structure's scattering of electrons and phonons. The effect of temperature on silver nanoparticles and AgNW is already studied experimentally and theoretically \cite{TempDepend_Ag_nanoparticle, AgNW_PhononScatt_SciRep, AgNW_thermal_effect, AgNW_phonon_tempDepend2022}; therefore, it is worth studying the effect of temperature on the silver nanowire-metamaterial to observe and understand how the system responds and leverages the plasmonic damping effects.\\
We first experimentally measured the spectroscopic properties of the HMM and HMM-quantum emitter coupled system and analyzed the results. Experimental data showed a non-monotonicity in the absorption and emission of interaction properties, which is directly related to the plasmon-damping. To understand that, we have theoretically investigated the effective damping mechanisms for AgNW from previous reports and applied them to numerically calculate spectral properties using Green's functions. We have performed numerical simulations using the finite difference time domain (FDTD) method and computed the scattering response of the HMM system in the temperature range of interest. The effect of temperature was invoked in the simulation by applying appropriate permittivities to the constituting materials of the HMM structure.\\
This paper is structured as follows: we first briefly describe the experimental and theoretical methods applied to this work, and then we discuss the experimental results and corresponding analyses. Then, to support and explain the data, we bring in the numerical simulation results and corresponding theoretical analyses. All relevant techniques are discussed in detail in the appendices.

\begin{figure*}
    \centering
    \includegraphics[width=1.0\textwidth]{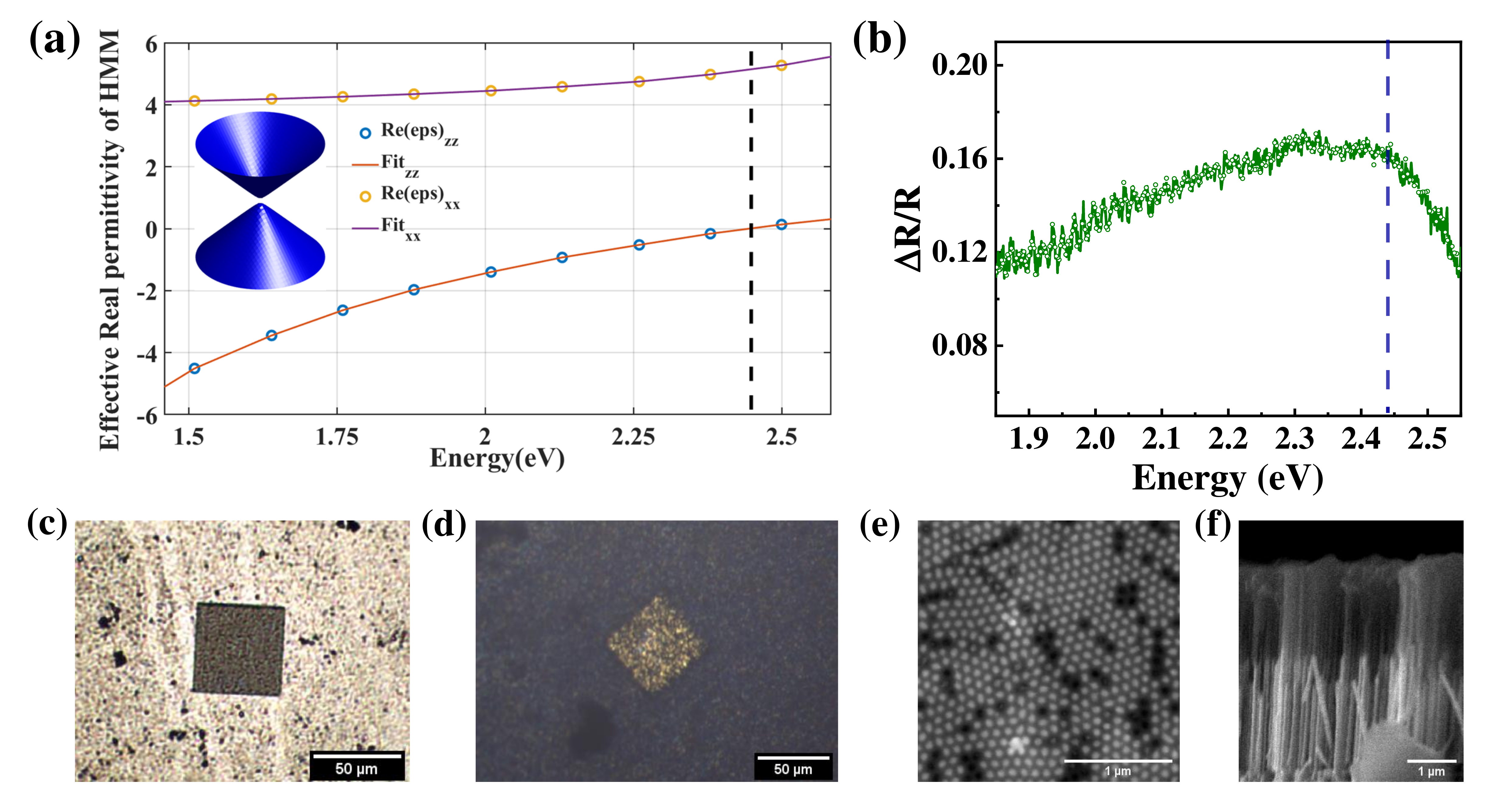}
    \caption{\textbf{(a)} HMM effective real permittivity along the nanowire growth direction (mentioned as $\epsilon_{zz}$) and along the plane of HMM (mentioned as $\epsilon_{xx}$); the dashed line shows the position of $\epsilon_{zz}$ crossing zero. The inset shows the schematic hyperboloid band structure for the available photonic density of states. \textbf{(b)} Differential reflection spectra of the HMM under broadband illumination to measure its absorption. The marked dashed line shows the OTT point corresponding to the dashed line shown in (a). \textbf{(c) \& (d)} Optical microscope image of FIB-etched HMM sample (square-shaped area); (c) before depositing the dye and (d) after deposition of the dye. \textbf{(e)} SEM image of the FIB-etched HMM sample; AgNWs grown in triangular lattice array. The darker sites are defects where AgNWs are undergrown. \textbf{(f)} Cross-section SEM image of the HMM shows the growth of AgNWs along the pores of porous Al$_2$O$_3$ structure.}
    \label{fig: Reeps_OTT_SEM}
\end{figure*}

\section{Methods}
\subsection{Experimental Methods}
HMM is fabricated using the chemical electro-deposition method of AgNWs within porous Al$_2$O$_3$ membrane in a solution of silver bromide (AgBr). Porous Al$_2$O$_3$ is obtained by two-step anodization of a high-purity Aluminum sheet. The filling fraction of AgNWs within the porous Al$_2$O$_3$ matrix is chosen by setting the pre-determined anodization voltage. After the deposition of AgNW, the final HMM sample is obtained, followed by a Focused Ion Beam (FIB) etching to remove the top metal layer \cite{HMM_SRK, HMM_RKY, HMM_Harsha}. The AgNW array is formed in a triangular lattice structure (Details of HMM fabrication are provided in Appendix \ref{append:HMM_fab}).  $10$ $\mu$L of $1.24$ $\mu$M/mL Rhodamine 6G (Rh6G) dye solution in methanol was spread over the FIB etched HMM interface, separated by a non-interacting 10 nm thick polymer spacer layer. The dye has the spontaneous emission transition frequency $\sim 2.25$ eV in free space, i.e., after the OTT of the HMM at the lower frequency side, where it obtains its hyperbolic behavior. An identical sample was prepared with dye solution deposited over a Si-SiO$_2$ substrate for control measurements. The HMM-dye sample was attached over a similar Si-SiO$_2$ substrate to hold and support it over the microscope stage. Photoluminescence measurements in frequency and time space (PL and TRPL) were performed with a confocal microscope (WiTec Alpha 300) with a 50$\times$ magnification 0.55 N.A. objective lens, connected to the CCD for PL signal collections and APD for time-correlated single-photon counting (TCSPC), i.e., TRPL measurements. Samples were placed inside the vacuum chamber of the MicrostatHe cryostat (Oxford Instruments) for the measurement from 4K to RT. A white-light lamp source and a pulsed-to-CW convertible 405 nm emission diode laser (PicoQuant PDL 800-D) were used for the broadband and the PL and TRPL measurements, respectively.\\
\subsection{Theoretical Methods}
To suffice experimental observations w.r.t the temperature, analytical models of the surface plasmon damping in an AgNW were ushered. Effective damping parameters have been included in complex material permittivities and applied to numerical simulations to produce the effect of temperature. Simulations were performed using the FDTD method with a commercially available software package (\textcopyright FDTD, Lumerical inc. \cite{Ansys}).
We have used HMM made of a periodic array of AgNW in a triangular lattice structure; the array is embedded in a porous Al$_2$O$_3$ matrix. Figure \ref{fig:HMM_schematic_paper2} shows the schematic of the model of HMM made of AgNW array within the porous Al$_2$O$_3$ block with quantum emitter molecules sparsely dispersed over the HMM interface for the light-matter interaction. The effect of temperature is seen as the enhanced noise in the emission spectra of the coupled system when the system temperature is increased (from blue to red).
The temperature dependence of the surface plasmon damping of HMM is analyzed following the effective damping channels of AgNW due to its electron-phonon scatterings and the significantly increased contribution of phonon-phonon scatterings, which is otherwise negligible in bulk metals. We have followed Matthiessen's rule and Bloch-Gr\"{u}neisen theory \cite{AgNW_PhononScatt_SciRep, Bloch-Gruneisen_Book, Bloch-Gruneisen_Data}, and electron-phonon interaction and phonon anharmonic interaction (Umklapp scattering) \cite{AgNW_phonon_tempDepend2022, TempDepend_PL_Ag, TempDepend_Ag_phonon, Umklapp_scatt_1, Umklapp_scatt_2, Umklapp_scatt_3, Umklapp_3} for necessary calculations.
\section{Experimental Results}
\begin{figure*}
    \centering
    \includegraphics[width=0.95\textwidth]{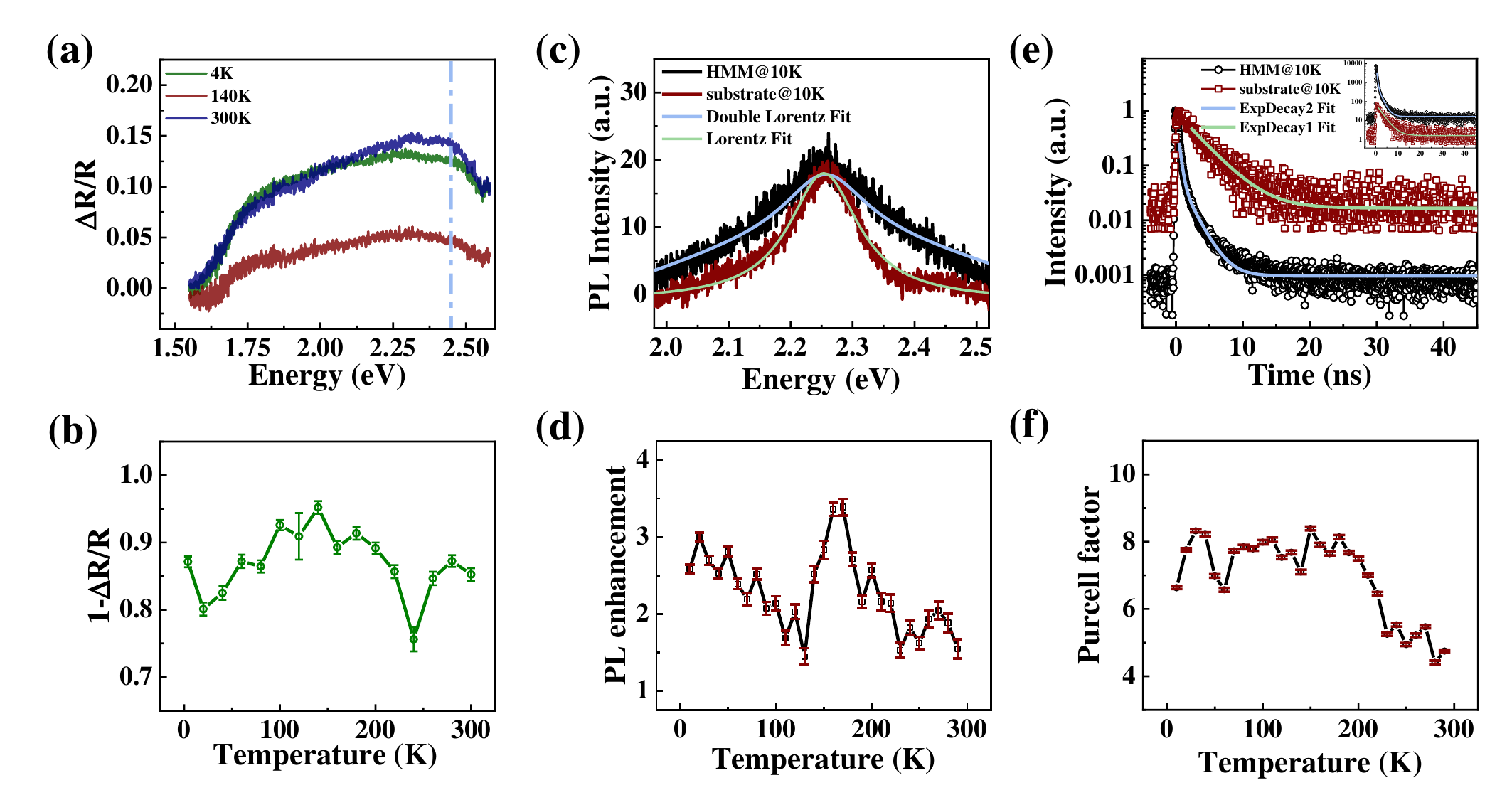}
    \caption{\textbf{(a)} 
    Differential reflection spectra of the HMM are shown at selected temperatures, 4K, 300K, and 140K. The dashed line around 2.43 eV marks the OTT. \textbf{(b)} Measured absorption parameter $\left(1-\frac{\Delta R}{R}\right)$ obtained from the differential reflection measurements, and their dependence w.r.t the temperature around the emitter transition frequency. \textbf{(c)} Photoluminescence (PL) scattering intensity plots for the dye-coupled HMM system and the dye over the substrate. \textbf{(d)} Corresponding PL enhancement for spontaneous emission of the dye-coupled HMM system over the temperature range. \textbf{(e)} shows the normalized spontaneous emission decay spectra. The inset shows the same plots with unnormalized intensities. \textbf{(f)} The analyzed Purcell factor for the spontaneous emission decay rates analyzed w.r.t the temperature.} 
    \label{fig: LT_absorb_PL}
\end{figure*}
The AgNW-HMM structure was characterized with scanning electron microscopy (SEM) imaging from the top part after the FIB etching and from the cross-section side of the HMM. Corresponding SEM images are shown in Fig.\ref{fig: Reeps_OTT_SEM}.(e) and (f). The FIB-etched samples were used for all the optical measurements, with and without depositing the dye molecules over them. Differential reflection spectroscopy is a useful tool to measure the absorption by thin films \cite{paper2_Diff_Reflect_1, paper2_Diff_Reflect_2, paper2_Diff_Reflect_3}. The HMM sample is a thin film compared to the thickness of the substrate (Si-SiO$_2$ substrate), which is evident from the cross-section SEM image [Fig.\ref{fig: Reeps_OTT_SEM}.(f)]. Examination of the HMM optical properties using the differential reflection spectroscopy is already been verified \cite{HMM_Harsha}, and it can efficiently estimate the OTT point for the HMM, which is predicted by its effective real permittivity value $\epsilon_{zz}$ along the nanowire axes direction crossing zero. Figure \ref{fig: Reeps_OTT_SEM}.(a) shows effective real permittivities of the HMM with filling fraction $f=0.2$, calculated from the Maxwell-Garnett effective medium theory (EMT) \cite{EMT_MG_1, EMT_MG_2, EMT_MG_3, EMT_MG_4}, along the nanowire growth direction ($\epsilon_{zz}$) and along the plane of HMM ($\epsilon_{xx}$ or $\epsilon_{yy}$). The dashed line marks where $\epsilon_{zz}$ crosses zero. After that point, for all the negative values of $\epsilon_{zz}$, the available states follow a hyperboloid isofrequency surface [inset of Fig.\ref{fig: Reeps_OTT_SEM}.(a)]. Figure \ref{fig: Reeps_OTT_SEM}.(b) shows the differential reflection spectra of the HMM with the filling fraction $f=0.2$. The observed scattering coefficient of the differential reflection, i.e., $\frac{\Delta R}{R}$ is around $2.43$ eV, and corresponds to the absorption property of the metamaterial starting with hyperbolic dispersion, which experimentally validates the OTT as predicted close to $2.43$ eV in Fig.\ref{fig: Reeps_OTT_SEM}.(a). Figure \ref{fig: Reeps_OTT_SEM}.(c) and (d) show the optical microscope images of the FIB-etched HMM sample before and after dye deposition. The square area is the HMM, where the top silver layer is etched out by FIB milling. Figure \ref{fig: Reeps_OTT_SEM}.(e)\&(f) present the SEM images of FIB-etched HMM and its cross-section to show the AgNW growth inside the porous Al$_2$O$_3$ matrix.

Figure \ref{fig: LT_absorb_PL}.(a) presents the differential reflection spectra of HMM under broadband planewave illumination over the temperature range 4K - 300K. Frequency dispersion data were shown at the lowest and highest temperatures and at 140K, where the absorption is found to be maximum. Hyperbolic transition around the OTT point is seen for all these plots (marked with a light blue dashed line near 
2.43 eV). $\frac{\Delta R}{R}$ presents the scattering response of the HMM thin film extracted from the total responses of the thin film-substrate system. Figure \ref{fig: LT_absorb_PL}.(b) plots the corresponding absorption parameter $\left(1-\frac {\Delta R}{R}\right)$ of the HMM thin film around the transition frequency of Rh6G emitters.
\\
Next, spectroscopic data for spontaneous emissions were taken from HMM with Rh6G dye molecules dispersed over it, as the sample device is shown in Fig.\ref{fig: Reeps_OTT_SEM}.(d). Experimental observations were made over several sample devices, and results were averaged over their corresponding data sets. Frequency-dependent photoluminescence (PL) spectra of spontaneous emissions collected from the HMM-dye system and the dye-substrate system for control measurements are shown in Fig.\ref{fig: LT_absorb_PL}.(c) at 10K temperature.
We analyzed PL enhancement for each temperature and plotted in Fig.\ref{fig: LT_absorb_PL}.(d). We presented the spontaneous emission decay spectra of the coupled system and the control system in Fig.\ref{fig: LT_absorb_PL}.(e). Spectrum from the HMM-dye system shows a bi-exponential decay with time, which is fitted by Eq.\ref{eq: biexponent},
\begin{equation}\label{eq: biexponent}
    \mathcal{A}(t)=\mathcal{A}_1e^{-\frac{t_1}{\tau_1}}+\mathcal{A}_2e^{-\frac{t_2}{\tau_2}}
\end{equation}

The decay rates of HMM modes and the free-space spontaneous emission decay rate of Rh6G are much different in order of magnitude; thus, the coupled system is expected to show a bi-exponential decay due to the availability of two decay paths \cite{RKY_RTSinglePhoton_Plasmonic}. These two decay paths also affect corresponding linewidths of the emission spectra of the coupled system; the slower path has a decay rate comparable to the free-space decay rate of the emitter, and the cavity-coupled path has a Purcell-enhanced faster decay rate. Hence, corresponding frequency spectra (PL spectra) are fitted with a double Lorentzian function (defined in Eq.\ref{eq: doubleLorentz}) of emission peak centered at $E_0$ and different linewidth parameters $\Gamma_1$ and $\Gamma_2$, respectively;
\begin{equation}\label{eq: doubleLorentz}
    \mathcal{F}=A_1\frac{\frac{\Gamma_1}{2}}{(E-E_0)^2+\left(\frac{\Gamma_1}{2}\right)^2}+A_2\frac{\frac{\Gamma_2}{2}}{(E-E_0)^2+\left(\frac{\Gamma_2}{2}\right)^2}
\end{equation}

We have studied the HMM system employing differential reflectivity measurements in Fig.\ref{fig: LT_absorb_PL}.(a) and (b) to observe its absorption responses with temperature before the emitter coupling. From the differential reflectivity, one can measure the absorption of the incident electromagnetic field exciting the thin film system \cite{paper2_Diff_Reflect_1, paper2_Diff_Reflect_2, paper2_Diff_Reflect_3}, given the assumption that there is no measurable transmission signal available in the plasmonic HMM system due to metallic losses. Therefore, $\left(1-\frac{\Delta R}{R}\right)\propto\alpha$, the absorption coefficient,
\begin{equation}\label{eq:diff_reflct}
   1- \frac{\Delta R}{R}=\frac{4dn_1n_2}{n_3^2-n_1^2}\alpha_2
\end{equation}
where $d$ is the thin film thickness, $n_1$ and $n_3$ are the refractive indices of materials above and below the film, respectively, and $n_2$ is the refractive index of the thin film. Reflectivity $R$ from a surface can be expressed in terms of its complex dielectric function $\epsilon$ \cite{Metal_reflectivity}, which is connected to the extinction coefficient $k$, the refractive index $n$, and we have discussed in Appendix \ref{app:alpha_oscillator strength} that material absorption is proportional to the characteristic oscillator strengths in that material induced by the incident electromagnetic field \cite{Ag_OpticalDielectricFunc, Ag_abs1_plasmonicblackbody, Ag_abs2_sizedepend, Ag_abs3, DiffReflect_KKR, RLoudon_propagation_EMwave}; Eq.(\ref{eq:SI_alpha_drude}) and Eq.(\ref{eq:SI_alpha_f}) find that the absorption coefficient of the material is proportional to the material absorption oscillator strength. An HMM thin film made of AgNWs has individual surface plasmons in the nanowires. Therefore, these expressions validate the absorption study, which directly connects the plasmonic metamaterial mode properties in this regard.
The experimentally measured absorption parameter $\left(1-\frac{\Delta R}{R}\right)$ is plotted in Fig.\ref{fig: LT_absorb_PL}.(b), it shows a non-monotonic functional behavior w.r.t the temperature, and finds its maximum around 150K, along with two minima around 20K and 220K. 

Next, let us focus on the discussion of spectral data obtained from the spontaneous emission of the dye-coupled HMM system. Figure \ref{fig: LT_absorb_PL}.(c) and (d) present the PL data and corresponding enhancement factor over the temperature range. PL enhancement finds a functional behavior similar to that of the absorption parameter with temperature. Parameters for the absorption and the spontaneous emission intensity enhancement include the coupling to the external electromagnetic field and the mode dephasing information. Therefore, these two parameters provide complete information relevant to light-matter interaction studies. Figure \ref{fig: LT_absorb_PL}.(e) and (f) present the corresponding spontaneous emission decays and the Purcell factor for the dye-coupled HMM system. Unlike the PL enhancement, the Purcell factor considers the radiative decays only; thus, it is related to the enhancement factor, but not the same as that. We found that the temperature-dependent functional behavior of the Purcell factor is comparatively flatter than the other two parameters of absorption and PL studies.\\ 
Since the experimental data for the material absorption and the spontaneous emission of dye emitters coupled to the HMM indicate a non-monotonic functional behavior of relevant parameters with temperatures, we need to carefully look at the inherent physical properties of the metamaterial to understand its origin. It is known that material properties, significantly for metals, rely on external parameters like the temperature \cite{HMM_Harsha,jacob2012broadband, PurcellEffect_origin_FieldTheory}. In this work, we have a metallic constituent, AgNW, in the HMM, which has a significant dispersion of material properties both with frequencies and temperatures, and that controls the plasmonic properties of the HMM. Hence, we need to delve into the temperature dependency of the material dielectric constants of an individual AgNW, and its corresponding effect in the AgNW-array HMM.

\section{Theoretical Results}
The temperature controls the plasmonic damping, and plasmonic damping is incorporated with the complex dielectric constant of the metal \cite{TempDepend_PL_Ag, TempDepend_Ag_phonon, AuSPP_tempdepend_AVZayats}. Experimental results connecting Purcell enhancement with the dielectric constant of the metallic thin film are observed to vary with temperature \cite{TempDepend_Agfilm_Purcell}. Therefore, we need to explore the complex dielectric function of AgNW with temperature. The effective damping arising due to the different internal scattering mechanisms are discussed in Appendix \ref{app:AgNW_phonon_scattrates}.\\
The surface plasmon damping affects the imaginary part of the material permittivity \cite{TempDepend_PL_Ag},
\begin{equation}\label{eq:im_eps_Tempdepend}
    \epsilon_i(\omega,T)=\epsilon_i(\omega,T_0)\frac{\mathcal{G}(T)}{\mathcal{G}(T_0)}
\end{equation}
where $\mathcal{G}$ signifies the surface plasmon damping parameter and $T_0$ signifies the room temperature.
\begin{figure}
    \centering
    \includegraphics[width=0.45\textwidth]{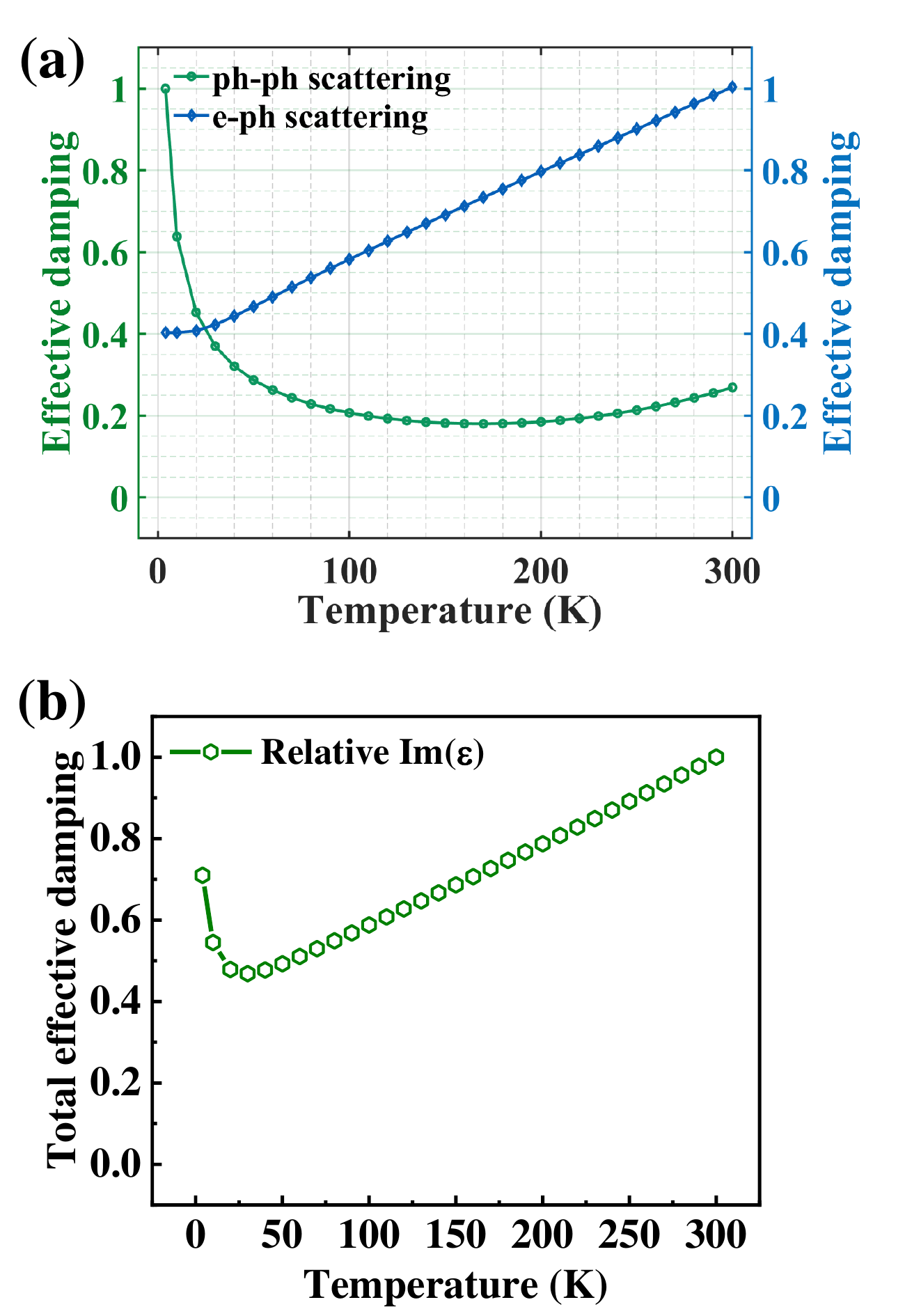}
    \caption{\textbf{(a)} Effective damping parameters of surface plasmons in AgNW due to electron-phonon (e-ph) and phonon-phonon (ph-ph) scatterings, and these need to be put in Eq.(\ref{eq:im_eps_Tempdepend}). \textbf{(b)} The total effective damping of the AgNW, which is also the total relative imaginary permittivity of the AgNW.}
    \label{fig: total_damping_ImEps}
\end{figure}
Figure \ref{fig: total_damping_ImEps}.(a) presents the effective surface plasmon damping parameter $\frac{\mathcal{G}(T)}{\mathcal{G}(T_0)}$ in an AgNW due to two scattering mechanisms, electron-phonon (e-ph) and phonon-phonon (ph-ph) scatterings, respectively. 
Only the e-ph scattering is significant in bulk metal; lowering the temperature freezes the phonon modes, thus decreasing the resistivity, which is almost vanishing around 10K. However, in AgNW, the nanoscale size of the metal causes an increase in the grain boundary and surface scatterings, which provides the residual resistivities at lower temperatures. The difference in temperature dependencies of the electrical and thermal resistivities in AgNW could be possible because of the discretization of electron and phonon modes, leading to a change in electron-phonon coupling strengths \cite{AgNW_PhononScatt_SciRep, metalnano_discretemode_sizeeffect}. With this change in sizes, the observed change in the Lorentz number and its variation with temperatures are studied in experiments and in theories for AgNWs \cite{AgNW_PhononScatt_SciRep, metalnano_ph_thermconduct}. It is found and reported that the phonon's contribution to the resistivity in AgNW is no longer negligible around lower temperatures, like 10K; the increased phonon contribution to the total resistivity of AgNW around 10K is around $30\%$ \cite{AgNW_PhononScatt_SciRep}. Eventually, this increases the significance of the phonon-phonon scattering and allows it to control the plasmonic damping in the AgNW.

\begin{figure*}
    \centering
    \includegraphics[width=0.99\linewidth]{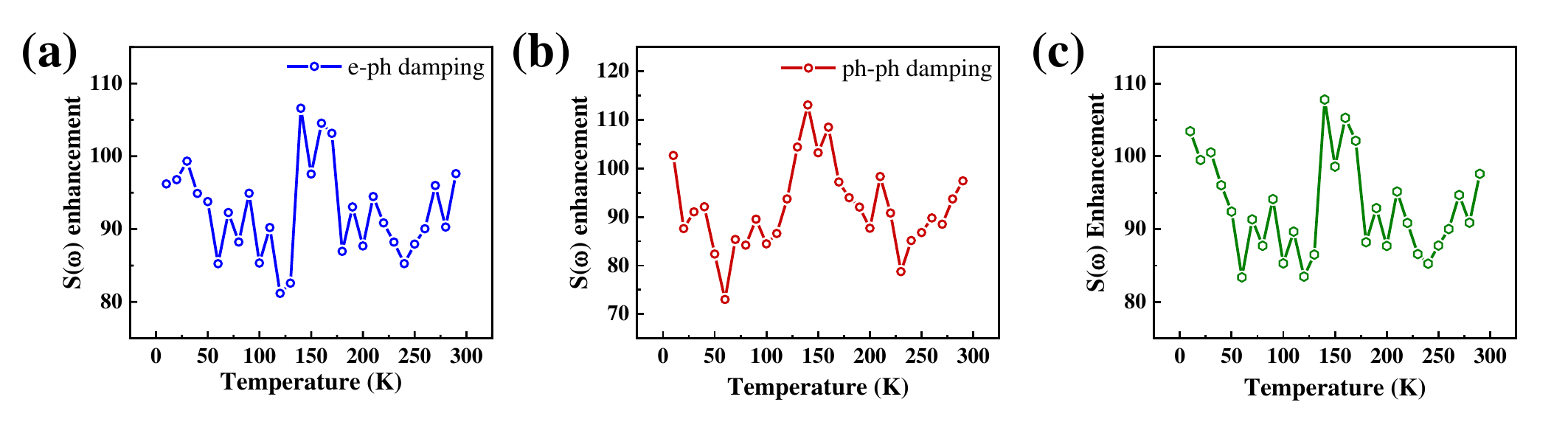}
    \caption{Numerical results for the scattering enhancement obtained from the HMM-dipole system and compared with the dipole-substrate system, for both damping models. \textbf{(a)} and \textbf{(b)} show the enhancement of the $S(\omega)$ function from the HMM-dipole system, and \textbf{(c)} shows corresponding total result.}
    \label{fig: Greens_total}
\end{figure*}
We took the experimentally observed $30\%$ proportion of increased phonon contributions around the lowest temperatures for a single AgNW \cite{AgNW_PhononScatt_SciRep}, where phonon modes are mostly frozen in bulk metal, to determine the total effective surface plasmon damping parameter, which was otherwise difficult to directly estimate in experiments \cite{EPI_scattrate_PRL}. Following the temperature dependence of Lorentz numbers for the AgNW, we added the effective damping constants to get the total one, which is also the relative imaginary dielectric constant of the AgNW, and plotted the result in Fig.\ref{fig: total_damping_ImEps}.(b). This result is very close to an experimental observation with silver thin films of the thickness comparable to the AgNW diameter in our work, and the excitation frequency used in that work is also the same as we worked \cite{total_Imeps_AgNW}. This explains that, unlike bulk silver, in AgNW, the effective plasmonic damping and the corresponding relative imaginary permittivities behave non-monotonically with temperature. Therefore, we need to apply these relative imaginary permittivity values to yield the scattering response of the HMM system with temperatures. We performed the finite difference time domain (FDTD) simulations in the model AgNW-HMM system using the commercially available FDTD package software \cite{Ansys} to get the scattering field responses of the system. We have used the room temperature data of complex permittivities of silver from Johnson \& Christy \cite{Johnson_Christy}. Temperature-dependent complex permittivities are determined following Eq.\ref{eq:im_eps_Tempdepend} \cite{TempDepend_PL_Ag}; the imaginary part of the complex permittivity $\epsilon$ is multiplied w.r.t the damping parameters of each scattering mechanism to get their values at respective temperatures.  
Simulations were performed with a dipole source placed in the near-field of the HMM interface. Details of the simulation setup are provided in the Appendix \ref{app: sim_setup}. Simulations were performed for both plasmon damping models.  Then we did calculations around the free-space dipole transition frequency of the dye $\simeq$ 2.25 eV for both damping models.\\
From the scattering field responses, we have defined the scattering Green's function of the HMM-dipole system, where the dye emitter is approximated with a dipole source \cite{OurAdomPaper}. Details are given in Appendix \ref{app: GreensFunction}. Here, in theoretical results, the frequency-dependent spectral function $S(\omega)$, defined to calculate the numerical scattering spectra of the system, is the most important function, which contains all the necessary information about the light-matter interaction.
\begin{gather}\label{eq: Greens}
S(\omega)=R_i(\omega)\chi(\omega),\\
R_i(\omega)=\frac{\sum_{j}G_{ji}^*(\mathbf{r}_D,\mathbf{r}_d,\omega)G_{ji}(\mathbf{r}_D,\mathbf{r}_d,\omega)}{\sum_{j}G_{ji}^{(0)*}(\mathbf{r}_D,\mathbf{r}_d,\omega)G_{ji}^{(0)}(\mathbf{r}_D,\mathbf{r}_d,\omega)},\; i,j=x,y,z,\\
\chi(\omega)=\frac{1}{[\{(\omega_{d}-\omega)T_{2}-X_{Re}\}^{2}+(1+X_{Im})^{2}]},\label{eq: denom_chi}\\
\begin{split}\label{eq: T1_T2_X}
    \mathrm{where,} &\quad X=X_{Re}+\mathtt{i}X_{Im}\\
&=\left(\frac{T_2}{2T_1}\right)\frac{G_{ii}(\mathbf{r}_d,\mathbf{r}_d,\omega)}{Im[G_{zz}^{(0)}(\mathbf{r}_d,\mathbf{r}_d,\omega)]},
\end{split}
\end{gather}
Equation \ref{eq: Greens} has the outcoupling function $R_i(\omega)$ for each dipole polarization direction $i$. It contains the information about the scattering and the influence of the metamaterial in the scattering defined by the Green's function. The denominator, $\chi(\omega)$, contains the information of the dipole emitter, expressed in Eq \ref{eq: denom_chi} - Eq.\ref{eq: T1_T2_X}. $T_1$ and $T_2$ are the longitudinal and transverse relaxation constants of the dipole source, considering it as a two-level system. Other details about the calculation are discussed in the Appendix \ref{app: GreensFunction}.\\
In this work, we have performed simulations for x and y polarizations of the dipole, respectively. These are along the HMM interface, assuming the interface is coplanar to the x-y plane. We have excluded the z-polarized dipole, since the coupling strength of this dipole-polarization at the HMM interface is negligible compared to the x and y components of the fields. The components of the local scattering field amplitude $E$ yield the coupling parameter $g=\vec{\mu}\cdot\vec{E}$ for the dipole. These field components are enhanced for a coupled system compared to free space. We found that for $E_{x,y}$, this enhanced amplitude value lies $\sim63$, and for $E_z$, it is $\sim1.75\times10^{-4}$. Therefore, z-polarized dipoles hardly couple to the HMM.  \\
Following Eq.\ref{eq: Greens} - Eq.\ref{eq: T1_T2_X}, we calculated the numerical scattering spectra, and compared results for the HMM-dipole and the dipole-substrate systems (Fig.\ref{fig: app_Sw} in Appendix \ref{app: GreensFunction}). Integrated intensities are compared for these two systems, and the ratio yields the enhancement factor. $S(\omega)$ enhancement is equivalent to the PL enhancement factor. Figure \ref{fig: Greens_total}.(a) and (b) present the $S(\omega)$ enhancement data with temperature for the e-ph and ph-ph damping, respectively. Figure \ref{fig: Greens_total}.(c) yields the summed result for these parameters following the results of Fig.\ref{fig: total_damping_ImEps}. Since the orientation of dye molecules was random in the experiment, and the dipole polarization would follow the excitation laser polarization, which lies along the HMM interface, therefore, all results shown in Fig.\ref{fig: Greens_total} are averaged with equal weight over the scattering responses for the x and y polarized dipole sources.\\
The unified numerical results predict the experimentally observed behaviors of relevant parameters with temperature very well, including the non-monotonicity. 

A comprehensive summary of the variation of factors for relevant observable parameters can be estimated. 
The variation of the effective damping, i.e., the total relative $Im(\epsilon)$ of AgNW lies within $1.43-2.13$ (Fig.\ref{fig: total_damping_ImEps}). Experimental results from Fig.\ref{fig: LT_absorb_PL} show a variation within a range of $1.023-2.34$; numerically simulated parameters also support that, the corresponding variation of factors is found to lie in a range of $1.06-1.29$, which is pretty close to the above-stated ranges of variation. This implies that the damping for surface plasmons within an AgNW-HMM lies much below that of individual AgNWs, which could be a significant gain in fabricating such metamaterials using lossy plasmonic materials.
Numerical results in Fig.\ref{fig: Greens_total} yield proximal results to the corresponding absorption and PL enhancement functions with temperature. The predicted maxima lie around 150K for both experiments and theories.
\section{Discussions}
To summarize, we thoroughly studied the temperature-dependent light-matter interaction of HMM made of AgNW-array embedded in Al$_2$O$_3$ matrix, starting from cryogenic temperature to room temperature. 
The collective array effect of surface plasmons from individual AgNWs provides cavity mode properties and ultrasmall mode volume. Here, we studied the light-matter interaction of the HMM-cavity properties by the absorption and the spontaneous emission from a weakly coupled system with dye emitters.  Experimental observations found a non-monotonous behavior with temperature, mostly showing maxima around 150K. Surprisingly, the Debye temperature for AgNW is 151K, and that of the bulk silver is 225K \cite{AgNW_PhononScatt_SciRep}. To understand the observed effects, we calculated numerical results following scattering Green's functions. We have simulated the model HMM-emitter system and the control system using the FDTD method; there, we incorporated the effect of temperature through the imaginary part of the complex permittivity of silver. We used two effective damping parameters from the analytical models of surface plasmon damping in AgNW, coming from electron-phonon and phonon-phonon scattering mechanisms. These two damping mechanisms are reportedly prevalent in AgNW \cite{AgNW_PhononScatt_SciRep}. We have chosen Rh6G dye molecules as quantum emitters because they don't have any emission mode dispersion w.r.t temperature; hence, only HMM mode's behavior could be captured. Next, we calculated numerical results of HMM scattering for the dipole excitation and found similar non-monotonic behaviors for the relevant, equivalent spectral enhancement parameters. We have analyzed and presented numerical results from both damping models separately and, in the end, added them to yield the complete result. We find that the effect of surface plasmon damping of AgNW is not affecting the AgNW-HMM to that extent, indicating the gain of metamaterial engineering made of lossy plasmonic materials. It is important to note that, in experiments and theories, the PL enhancement and $S(\omega)$ enhancement factors show a \lq{roughness}\rq; we found this only because of the transverse relaxation constant $T_2$. We have used $T_2$ values obtained from the control measurements and applied them in numerical calculations, and that confirms the inference. We found that Purcell factors are relatively flat in a broad region in the range 80K - 220 K. Purcell data includes the information of radiative lifetime $T_1$, and thus, it is expected to show differences from PL and absorption data. In fact, our results show that the radiative coupling of the light-matter interaction of the HMM system is quite robust against plasmonic damping losses.\\
It should be noted that we have emphasized the functional behaviors of numerical results, not the absolute values. In the experiment, we studied the collective emission of the HMM-emitter ensemble system. An identical emitter ensemble was deposited over the substrate for control measurements. In the simulation, however, we work with a single dipole emitter source. Therefore, the absolute values for experiments and theoretical results can't be compared; we can only compare the functional behavior, and the corresponding ratios are relevant.
\section{Conclusion}
Plasmonic metamaterials are a system that can support non-Hermitian phenomena due to their complex dielectric constants \cite{nonHermititan_review, nonHermitian_Optica}. 
We have used HMMs coupled with active gain media, i.e., quantum emitters, to balance loss and gain. In this work, we have presented the HMM system with and without the coupling to a gain medium and shown its strong absorption and spontaneous emission enhancement properties, respectively. We showed that the imaginary dielectric constant of silver nanowires plays a crucial role in determining the absorption and emission properties, which vary non-monotonously w.r.t temperature due to their internal scattering mechanisms. This would help determine and control extremely anisotropic metamaterial properties for imaging, broadband absorbing, sensing, chirality, etc. \cite{nonHermitian_SPP_anisotropy, nonHermitian_broadband_metamaterial, nonHermitian_OTT_Alu_PRL}, where non-Hermitian physics is prevalently present and holds the key in two ways: tuning w.r.t the temperature according to the need of applications and knowing the electromagnetic interaction behavior to apply at a fixed temperature.
Thus, the thorough study about the AgNW-HMM properties for light-matter interactions over a wide range of temperatures that we conducted to understand the key parameters tuning and controlling HMM cavity mode properties could help in these applications. Further research could be conducted based on the parity-time symmetry in such metamaterial systems \cite{nonHermitian_PTsymmetry_review, nonHermitian_broadband_metamaterial} for fundamental studies and engineering applications.
\section{Acknowledgement}
J.K.B. acknowledges funding from the Anusandhan National Research Foundation (ANRF), India, through grant number CRG/2021/003026 and DST, FIST grant. The authors thank Prof. Girish S. Agarwal, Texas A \& M University, USA, for his valuable suggestions and feedback on the theoretical modeling and calculations. A.N. thanks the Advanced Facility for Microscopy and Microanalysis (AFMM),
Indian Institute of Science, Bangalore, for access to SEM
and FIB facilities. A.N. thanks Harshavardhan R.K.for useful discussions. 
\appendix
\section{HMM Fabrication}\label{append:HMM_fab}
AgNW-based-HMM is fabricated by growing AgNWs within porous Al$_2$O$_3$ matrix. Al$_2$O$_3$ matrix is grown over high-pure Aluminum sheets procured from commercial supplier (Alfa Aesar), by two-step anodization. All the fabrication steps including sonication cleaning, electro-polishing, preparing for anodization with HPA acrylic coating in the polished-unpolished boundary to suppress surface current flow etc. have been carried out carefully following the methods as mentioned in \cite{HMM_Harsha,HMM_RKY} thoroughly. The filling fraction $f$, defined for the triangular lattice unit cell as,
\begin{equation}
    f=\frac{\pi}{2\sqrt{3}}\left(\frac{d}{a}\right)^2
\end{equation}
where $d$ is the nanowire diameter and $a$ is the separation between adjacent nanowire separation,
is determined by the anodization voltage, which is set to 40V to get $f=0.2$. Two step anodization for 14 hours and 2 hours, respectively, are conducted with maintaining a stable set voltage at 40 V and set current at 3 mA. After growing first oxide layer for 14 hours over the metal sheet, that layer is removed with dilute Chromic acid solution (2\% w/V) for 8 hours. Cleaned metal sheet is set for second anodization following previous steps for next 2 hours, which provides regularized arrays of pores.\\
After 2nd anodization, metal sheet from the bottom of the grown Al$_2$O$_3$ layer is etched and removed with mixed solution of CuCl$_2$-HCl. Barrier layer of Al$_2$O$_3$ is removed with Orthophosphoric acid solution (6\% v/V). This step ensures pore opening through the Al$_2$O$_3$ thin film; after confirming with AFM imaging about pore opening, a 5-10 nm thin gold film is sputtered over one side of Al$_2$O$_3$ thin film to act as a nano-electrode, and placed within AgBr solution with NaSO$_3$ and Na$_2$SO$_3$ to electro-chemically deposit Ag$^+$ ions through the pores to grow AgNWs inside that, 
thus preparing the HMM thin film. During the deposition, a layer of silver gets deposited over the other side of gold nano-electrode. That layer is etched and removed with FIB milling to finally receive the active HMM sample. The SEM images shown in Fig.\ref{fig: Reeps_OTT_SEM}.(e) find the AgNW diameter $\simeq60$ nm and array periodicity $\simeq120$ nm.
\section{Simulation setup} \label{app: sim_setup}
We have performed finite difference time domain (FDTD) simulations on a model AgNW-HMM system. The 3D simulation box contained AgNWs of diameter 60 nm and periodicity 120 nm in a triangular lattice structure, as obtained from the experiment; periodic boundaries were applied along the plane of the HMM interface, and perfectly matched layer boundaries along the normal directions to the plane. The alumina host medium had a permittivity of 2.56 applied in the numerical calculations. AgNW permittivities are adopted from Johnson and Christy's data \cite{Johnson_Christy}, and that was modified for each temperature. Simulations were performed with a dipole source kept at the near-field of the HMM interface for the near-field interaction. Corresponding scattering responses were recorded with a time monitor. Scattering field responses were normalized w.r.t the corresponding free-space results to simplify them, remove all dependencies like the dipole moment, excitation power, etc., and present results in arbitrary units. These field responses defined the scattering Green's function and calculated the relevant theoretical results.
\section{Numerical calculations of the spectral functions}\label{app: GreensFunction}
We have used the scattering Green's function to calculate the scattering spectra of the HMM excited by a dipole source at the near field. We need to out-couple radiation from the dipole position, $r_d$, to the detector position, $\textbf{r}_D$. This out-coupling is obtained via the Green's function, $\textbf{G}(\textbf{r}_D,\textbf{r}_d,\omega)$, using Maxwell's equations
\begin{equation} \label{eq: Greens_define}
    \textbf{E}(\textbf{r}_D,\omega) \longrightarrow \int \textbf{G}(\textbf{r}_D,\textbf{r}^\prime_d,\omega)\cdot \frac{\textbf{P}(\textbf{r}^\prime,\omega)}{\epsilon_0} d^3r^\prime
\end{equation}
\begin{figure}[t]
    \centering
    \includegraphics[width=0.45\textwidth]{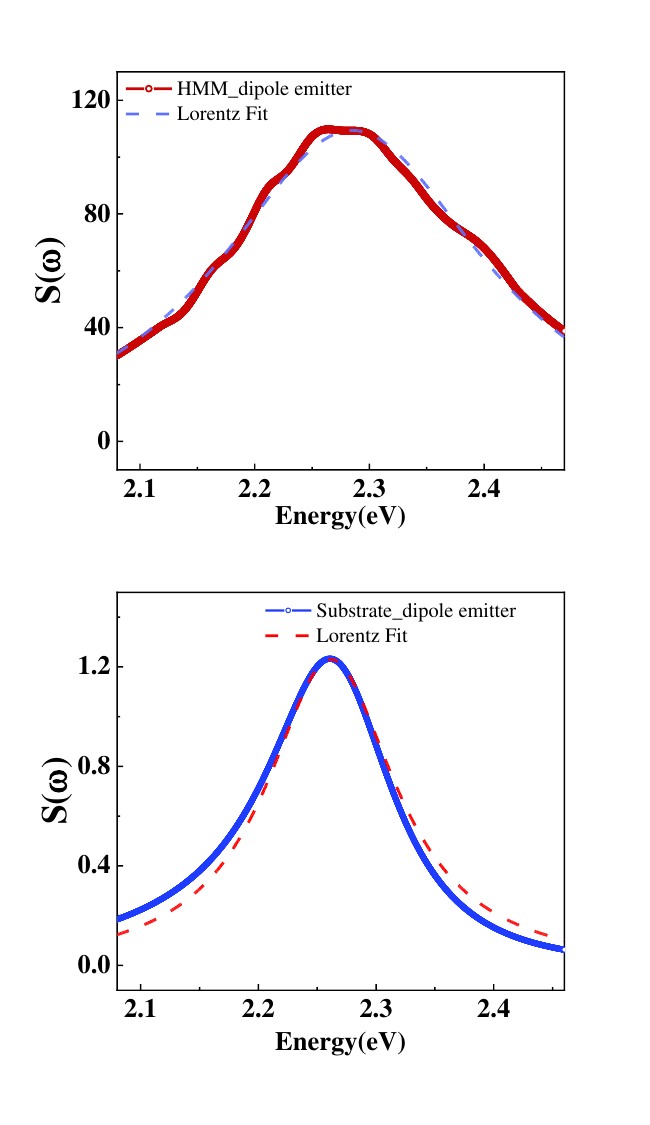}
    \caption{Scattering spectra calculated using Green's function for the HMM-dipole and substrate-dipole system.}
    \label{fig: app_Sw}
\end{figure}
\begin{figure*}
    \includegraphics[width=0.98\textwidth]{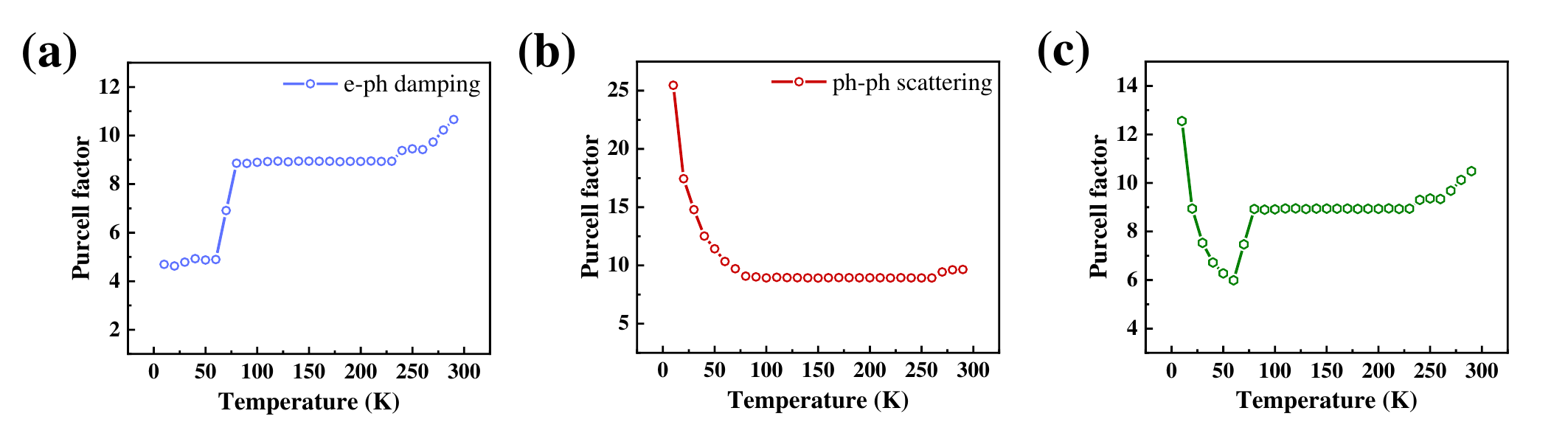}
    \caption{Theoretical Purcell factor results; \textbf{(a)} and \textbf{(b)} show the Purcell factor from the ratio of results of Eq.\ref{eq: appPurcell_ImY} for HMM-dipole and dipole-substrate systems, for both damping models. \textbf{(c)} shows the corresponding summed result.}\label{fig: apeend_Purcell_theory}
\end{figure*}
Here, the scattering field $\textbf{E}(\textbf{r}_D,\omega)$ is connected to the scattering environment of the dipole at source position, which is determined by the material polarizability $\textbf{P}(\textbf{r},\omega)$.
It is convenient to normalize the resulting spectrum w.r.t the peak value of the spectrum in free space $S^{(0)}(\omega_d)$. The normalized spectrum is obtained as,
\begin{gather}\label{eq: appGreens}
S(\omega)=R_i(\omega)\chi(\omega),\\
R_i(\omega)=\frac{\sum_{j}G_{ji}^*(\mathbf{r}_D,\mathbf{r}_d,\omega)G_{ji}(\mathbf{r}_D,\mathbf{r}_d,\omega)}{\sum_{j}G_{ji}^{(0)*}(\mathbf{r}_D,\mathbf{r}_d,\omega)G_{ji}^{(0)}(\mathbf{r}_D,\mathbf{r}_d,\omega)},\; i,j=x,y,z,\\
\chi(\omega)=\frac{1}{[\{(\omega_{d}-\omega)T_{2}-X_{Re}\}^{2}+(1+X_{Im})^{2}]},\\
\begin{split}
    \mathrm{where,} &\quad X=X_{Re}+\mathtt{i}X_{Im}\\
&=\left(\frac{T_2}{2T_1}\right)\frac{G_{ii}(\mathbf{r}_d,\mathbf{r}_d,\omega)}{Im[G_{zz}^{(0)}(\mathbf{r}_d,\mathbf{r}_d,\omega)]},
\end{split}
\\\mathrm{and},\quad Y=Y_{Re}+\mathtt{i}Y_{Im}=\frac{G_{ii}(\mathbf{r}_d,\mathbf{r}_d,\omega)}{Im[G_{zz}^{(0)}(\mathbf{r}_d,\mathbf{r}_d,\omega)]}\label{eq: appPurcell_ImY}
\end{gather}
The incoherent part of the free-space scattering spectrum of the dipole source is given by \cite{OurAdomPaper},
\begin{equation} \label{eq: S0}
    S^{(0)}(\omega) \propto g^2 (\frac{2T_1}{T_2}-1)\frac{1}{\pi}\frac{1}{\frac{1}{T_2^2}+(\omega - \omega_d )^2}
\end{equation}
which includes the coupling parameter $g$, longitudinal and transverse relaxation constants $T_1$ and $T_2$, respectively. The modified scattering response that incorporates the Green's function, defined by Eq.\ref{eq: Greens_define}, carries the information about the metamaterial's effect, gets added to the denominator of Eq.\ref{eq: S0}, and yields the results as seen in Eq.\ref{eq: appGreens} - Eq.\ref{eq: appPurcell_ImY}. $T_1$ values were obtained from the Rh6G emitter spontaneous emission decay rates in free space, and $T_2$ corresponds to the free space PL emission linewidth, measured in meV, and 1 meV corresponds to 0.66 ps \cite{OurAdomPaper}. In this work, $T_1$ remained $\sim2-3$ ns, and $T_2\sim5.1\times10^{-15}$s, therefore, $\left(\frac{T_2}{2T_1}\right)\sim1.2\times10^{-6}$. \\
We can obtain the theoretical Purcell factor by taking the ratio between the imaginary part of $Y$ results in Eq.\ref{eq: appPurcell_ImY} for the HMM-dipole system and the dipole-substrate system. Figure \ref{fig: apeend_Purcell_theory} shows the calculated Purcell factor results for both damping models. The summed result in Fig.\ref{fig: apeend_Purcell_theory}.(c) finds a similar flat region $\sim 60$K$-230$K as observed for the experimentally observed Purcell factor. These flat regions may emphasize that the radiative coupling of the HMM system with electromagnetic modes is more robust against external effects that could degrade it. However, the disagreement after $\sim230$K could be due to increasing surface and defect scattering effects, which might not permit a better coupling between the HMM and the emitter.
\section{Relation of Absorption coefficient to Absorption Oscillator strength of HMM}\label{app:alpha_oscillator strength}
Material absorption properties of HMM thin film would contribute in its photonic properties in light-matter interaction, where plasmons of constituent AgNWs would assume a key role; hence, a thorough understanding of absorption, oscillator strength and their connection is necessary to comprehend \cite{Ag_OpticalDielectricFunc, Ag_abs1_plasmonicblackbody, Ag_abs2_sizedepend, Ag_abs3, DiffReflect_KKR, RLoudon_propagation_EMwave}. 
Differential Reflection Spectroscopy measured absorption of the incident field in the HMM thin film; reflectivity of a surface is related to the material complex permittivity $\epsilon$ as \cite{Metal_reflectivity},
\begin{equation}\label{eq:SI_reflectivity}
    R=\left|\frac{\sqrt{\epsilon}-1}{\sqrt{\epsilon}+1}\right|^2
\end{equation}
Where $\epsilon=\epsilon_r+i\epsilon_i$; $\epsilon_r$ and $\epsilon_i$ can be expressed in terms of plasma frequency $\omega_p$,
\begin{gather}\label{eq:SI_er}
    \epsilon_r=\epsilon_\infty-\frac{\omega_p^2}{\omega^2+\Gamma_{ep}^2},\\
    \epsilon_i=\frac{\omega_p^2\Gamma_{ep}}{\omega(\omega^2+\Gamma_{ep}^2)}\label{eq:SI_ei},\\
    \omega_p=\sqrt{\frac{Ne^2}{\epsilon_0m}}
\end{gather}
$\epsilon_\infty$ is background dielectric constant due to core ions of metal, $\Gamma_{ep}$ is electron-phonon scattering rate, $N$ is conduction electron number density in the material, $e$ is electron charge and $m$ is effective electron mass.\\
Material refractive index $n$ and extinction coefficient $k$ are related to $\epsilon$,
\begin{gather}\label{eq:SI_sqrt_e}
    \sqrt{\epsilon}=n-ik,\\
    \implies \epsilon_r+i\epsilon_i=n^2-k^2+2ink\label{eq:SI_2nk}
\end{gather}
Equation (\ref{eq:SI_ei}) is used in Eq.(\ref{eq:SI_2nk}) to get,
\begin{equation}
    k=\frac{\omega_p^2\Gamma_{ep}}{2n\omega(\omega^2+\Gamma_{ep}^2)}
\end{equation}
and extinction coefficient $k$ is related to absorption coefficient, $\alpha=\frac{\omega}{c}k$. Hence, absorption coefficient for the material can be defined as,
\begin{equation}\label{eq:SI_alpha_drude}
    \alpha=\frac{Ne^2\Gamma_{ep}}{2nc\epsilon_0m(\omega^2+\Gamma_{ep}^2)}
\end{equation}
Now, let us consider plasma oscillations of the plasmons of constituent AgNWs within HMM; we can define absorption oscillator strength for them \cite{EinsteinCoeff}.\\
Transition dipole moment $\mu$ is connected to absorption oscillator strength $f$ as,
\begin{equation}\label{eq:SI_mu_f}
    |\mu|^2_{12}=\frac{3}{2}\frac{\hbar e^2}{\omega_0m}f_{12}
\end{equation}
where $\omega_0$ is the oscillator transition frequency, and 1,2 are the lower and upper energy states involved in the transition, respectively. Absorption cross-section $\sigma$ can be expressed in terms Einstein B coefficient,
\begin{gather}\label{eq:SI_sigma_EinsteinB}
    \sigma=\frac{\hbar\omega_0B_{12}}{c},\\
    \text{and,}\;\;\; B_{12}=\frac{g_2}{g_1}\frac{\pi^2c^3}{\hbar\omega_0^3}A_{21}\label{eq:SI_Einstein_BA}
\end{gather}
$A_{21}$ is Einstein A coefficient, $g_1$ and $g_2$ are degeneracy factors. Absorption coefficient $\alpha'$ connects with $\sigma$ as $\alpha'=N'\sigma$, $N'$ is the oscillator number density. Hence, putting Eq.(\ref{eq:SI_Einstein_BA}) in Eq.(\ref{eq:SI_sigma_EinsteinB}) it yields,
\begin{gather*}
    \sigma=\hbar\omega_0\frac{g_2}{g_1}\frac{\pi^2c^3}{\hbar\omega_0^3c}A_{21},\\
    \text{and,}\;\;\; A_{21}=\frac{2\omega_0^3\mu^2_{21}}{3\epsilon_0 hc^3},\\
    \text{hence,}\;\;\; \sigma=\hbar\omega_0\frac{g_2}{g_1}\frac{\pi^2c^3}{\hbar\omega_0^3}\frac{2\omega_0^3\mu_{21}^2}{3\epsilon_0hc^4},\\
    =\frac{g_2}{g_1}\frac{2\omega_0\pi^2}{3\epsilon_0hc}\mu_{21}^2
\end{gather*}
Using Eq.(\ref{eq:SI_mu_f}) and the relation $f_{21}=\frac{g_1}{g_2}f_{12}$, final expression is obtained;
\begin{gather}\label{eq:SI_sigma_f}
    \sigma=\frac{\pi e^2}{2\epsilon_0mc}f_{12},\\
    \alpha'=\frac{N'\pi e^2}{2\epsilon_0mc}f_{12}\label{eq:SI_alpha_f}
\end{gather}
Equation (\ref{eq:SI_alpha_f}) connects absorption coefficient for the oscillators to the absorption oscillator strength $f_{12}$. Comparing Eq.(\ref{eq:SI_alpha_drude}) and Eq.(\ref{eq:SI_alpha_f}) it is evident that, $\alpha\propto f$. Thus, absorption result obtained from differential reflection spectroscopy of HMM thin films do estimate its absorption oscillator strength of plasmons of constituent AgNWs, which determines HMM cavity field strength and mode linewidth. Material permittivity $\epsilon$ used in Eq.(\ref{eq:SI_reflectivity} - \ref{eq:SI_2nk}) is also valid for effective permittivity obtained from Maxwell-Garnett Effective medium theory \cite{Ag_abs1_plasmonicblackbody}. 
\section{Scattering rates and effective Damping mechanism in AgNW}\label{app:AgNW_phonon_scattrates}
Effective damping of plasmons in AgNW would determine light-matter interactions, in case of composite metamaterials also.\\
Compared to bulk Ag, in AgNW strong electron-phonon interaction is present due to reduction in dimensions \cite{AgNW_PhononScatt_SciRep}; electron-phonon interaction determines the effect of temperature in electron resistivity in AgNW. This resistivity term is given by Bloch-Gr\"{u}neisen theory \cite{Bloch-Gruneisen_Data,Bloch-Gruneisen_Book},
\begin{gather}
    \rho=\rho_0+\rho_{el-ph},\\
    \rho_{el-ph}=\alpha_{el-ph}\left(\frac{T}{\theta}\right)^n\int_0^{\frac{\theta}{T}}\frac{x^n}{(e^x-1)(1-e^{-x})}dx
\end{gather}\label{eq:SI_Bloch_Gruneisen}
Where $\alpha_{el-ph}$ is electron-phonon coupling constant, $\theta$ is the Debye temperature of AgNW, $n=5$ chosen for non-magnetic metals. Thermal conductivity emerging from electronic conduction is given by Wiedemann-Franz law \cite{Wiedemann_Franz,Wiedemann_Franz_2},
\begin{equation}\label{eq:Wiedemann_Franz}
    k_e=\frac{L_0T}{\rho}
\end{equation}
$L_0$ is Lorentz number and $T$ is absolute temperature.\\
Apart from electronic conduction, thermal conductivity in metal is also dependent to lattice conductivity \cite{AgNW_phonon_tempDepend2022,Wiedemann_Franz_2}, which is independent of carrier concentration.\\
Lattice thermal conductivity depends on electron-phonon scattering dominant at lower temperatures, phonon-phonon Umklapp scattering dominant at higher temperatures \cite{Umklapp_scatt_1,Umklapp_scatt_2,Umklapp_scatt_3} and other contributions like surface boundary scattering, phonon structure scattering etc. In case of AgNW, total phonon scattering rate can be expressed as,
\begin{equation}\label{eq:SI_total_phonon}
    \gamma_T=\gamma_{el-ph}+\gamma_{Um}+\gamma_b
\end{equation}
$\gamma_{el-ph}$ is phonon scattering rate due to electron-phonon interaction, $\gamma_{Um}$ represents anharmonic phonon-phonon scattering, i.e., Umklapp process, and $\gamma_b$ emerges due to phonon scattering at surface boundaries. For AgNW, $\gamma_b$ is almost constant over temperature \cite{AgNW_phonon_tempDepend2022}, therefore, temperature dependent damping is controlled by $\gamma_{el-ph}$ and $\gamma_{Um}$.\\
It is assumed that acoustic phonon modes are responsible for thermal conduction in lattice; hence, electron-phonon scattering for acoustic phonon modes is given by \cite{EPI_scattrate_PRL},
\begin{gather}\label{eq:SI_EPI_scattrate}
    \gamma_{el-ph}=\frac{\sqrt{2\pi m^*}E^2_D}{(kT)^{\frac{3}{2}}gd_mv_s}\text{exp}\left(-\frac{m^*v_s^2}{2k_BT}\right)n(E_F)\omega,\\
    n(E_F)=\left(\frac{1}{3\pi^2}\right)\left(\frac{2m^*E_F}{\hbar^2}\right)^{\frac{3}{2}}\label{eq:SI_eConcentration_Fermi}
\end{gather}
$m^*$ is density-of-state effective mass of carrier in solid, $g$ is number of equivalent carrier pockets, $d_m$ is mass density of material, $v_s$ is the average sound velocity in the solid, $E_D$ is the deformation potential appearing from Deformation Potential Approximation since phonons cause a uniform strain in lattice; $n(E_F)$ is the carrier concentration with Fermi energy $E_F$. For AgNW, $m^*=0.4m_e$, $g=10$, $E_D=2.9$ eV and $E_F=3.1$ eV.\\
Anharmonic phonon-phonon scattering, i.e., Umklapp scattering is stronger in higher temperatures \cite{Umklapp_scatt_1, Umklapp_scatt_2, Umklapp_scatt_3, Umklapp_3}; corresponding scattering rate is given by,
\begin{equation}\label{eq:SI_Umklapp}
    \gamma_{Um}=\frac{\hbar\varsigma^2\omega^2T}{M_av_s\theta_D}e^{-\frac{\theta}{3T}}
\end{equation}
$\varsigma$ is Gr\"{u}neisen parameter = 2.29 for silver, $M_a$ is average mass of atom in lattice, $\theta_D$ is Debye temperature.\\
Lattice thermal conductivity is expressed as \cite{AgNW_phonon_tempDepend2022,latt_therm_cond},
\begin{gather}\label{eq:SI_latt_therm_cond}
    k_L=\frac{1}{3}C_vv_s^2\gamma_T,\\
    C_v\propto\left(\frac{T}{\theta_D}\right)\int_0^{\frac{\theta_D}{T}}\frac{x^2e^x}{(e^x-1)^2}dx \label{eq:SI_Cv}
\end{gather}
Therefore, it is easy to comprehend that effective damping of surface plasmons in AgNW is proportional to resistivity in Eq.(\ref{eq:Wiedemann_Franz}) and inverse of these scattering rates in Eq.(\ref{eq:SI_latt_therm_cond}). Hence, those effective damping parameters are plotted in Fig.\ref{fig: total_damping_ImEps}.(a), which would be used to calculate temperature effects on imaginary dielectric constant of silver in Eq.\ref{eq:im_eps_Tempdepend}.
\FloatBarrier
\bibliography{HMM2} 

\end{document}